\documentstyle[12pt]{article}

\newcommand{\om}{\omega}
\newcommand{\vp}{\varphi}

\begin{document}

\begin{center}

{\Large{\bf Critical Effects in Population Dynamics of Trapped
Bose-Einstein Condensates} \\ [5mm]

V.I. Yukalov$^1$, E.P. Yukalova$^2$, and V.S. Bagnato$^3$} \\ [3mm]

{\it $^1$Bogolubov Laboratory of Theoretical Physics\\
Joint Institute for Nuclear Research, Dubna 141980, Russia \\ [3mm]

$^2$Department of Computational Physics\\
Laboratory of Informational Technologies\\
Joint Institute for Nuclear Research, Dubna 141980, Russia \\ [3mm]

$^3$Instituto de Fisica de S\~ao Carlos, Universidade de S\~ao Paulo \\
Caixa Postal 369, S\~ao Carlos, S\~ao Paulo 13560-970, Brazil}

\end{center}

\vskip 3cm

\begin{abstract}

The population dynamics of a trapped Bose-Einstein condensate, subject to 
the action of an external field, is studied. This field produces a 
spatio-temporal modulation of the trapping potential with the frequency 
close to the transition frequency between the ground state and a higher 
energy level. For the evolution equations of fractional populations, a 
critical line is found. It is demonstrated that there exists a direct 
analogy between dynamical instability at this line and critical 
phenomena at a critical line of an averaged system. The related critical 
indices are calculated. The spatio-temporal evolution of atomic density is 
analyzed.

\end{abstract}

\newpage

\section{Introduction}

A dilute cloud of Bose-condensed atoms confined within a trapping potential
at low temperatures can be described by a wave function satisfying the
Gross-Pitaevskii equation [1,2]. This equation is nonlinear due to the 
atomic interactions through a local delta potential. The mathematical
structure of the equation is that of the nonlinear Schr\"odinger equation.
Stationary states of the latter, because of the confinement caused by a 
trapping potential, are restricted to discrete energy levels. These 
stationary eigenstates form a set of {\it nonlinear modes}, in analogy to 
linear modes that are solutions of a linear Schr\"odinger equation. The 
nonlinear modes can be called {\it coherent modes}, since the wave function 
of the Gross-Pitaevskii equation corresponds to a coherent state of 
Bose-condensed atoms. It is also possible to use the term {\it topological 
modes}, emphasizing that the wave functions related to different energy 
levels have different spatial topology.

This modal structure of the confined condensate states is in close analogy 
with {\it nonlinear optical modes} of nonlinear waveguide equations in 
optics [3,4]. And the methods of creating nonlinear coherent modes of trapped
Bose atoms [5-7] are also similar to those employed in optics, where one 
uses specially prepared initial conditions or invokes an action of external 
fields. Dynamics of fractional populations, characterizing the occupation
of coherent modes of Bose-condensed trapped atoms, displays as well many 
phenomena equivalent to those known in optics, for instance collapses, 
revivals, and Rabi-type oscillations. The nonlinear dynamics of processes 
coupling bosonic coherent modes have been studied in several papers 
considering the general case [5-7], antisymmetric modes [8,9], dipole 
topological modes in a two-component condensate [10], dark-soliton states 
[11], and vortex modes [12-19].

The aim of the present paper is threefold: First, we demonstrate the existence 
of a critical line for the population dynamics of Bose condensates, 
which is a kind of bifurcation line due to the nonlinearity of evolution 
equations. Second, we show that there is an intimate relation between the 
studied nonlinear dynamical system and the corresponding averaged system, 
so that the bifurcation line is an analog of the  critical line, and 
instabilities happening at the former are the counterparts of critical 
phenomena occurring at the latter. Third, we study the spatio-temporal 
evolution of atomic density under the action of a resonant external field.

\section{Critical Phenomena}

The time-dependent Gross-Pitaevskii equation has the form
\begin{equation}
\label{1}
i\hbar\;\frac{\partial\varphi}{\partial t} = \left [ \hat H(\varphi) +
\hat V \right ]\; \varphi\; , \qquad
\hat H(\varphi)= \; - \; \frac{\hbar^2}{2m_0}\; {\bf\nabla}^2 +
U({\bf r}) + AN|\varphi|^2 \; ,
\end{equation}
where $U({\bf r})$ is a trapping potential; $A\equiv4\pi\hbar^2a_s/m_0$; 
$a_s$ is a scattering length; $m_0$ is mass; and $N$ is the number of 
particles; $\hat V$ is a potential of external fields. The wave function 
$\varphi$ is normalized to unity, $||\varphi||=1$. The nonlinear 
topological modes are defined [5-7] as the eigenfunctions of the nonlinear 
Hamiltonian, that is, they are the solutions to the eigenproblem
$$
\hat H(\varphi_n)\varphi_n({\bf r}) =E_n\varphi_n({\bf r}) \; .
$$
It is worth stressing here the principal difference between the topological 
coherent modes and collective excitations. The former are self-consistent 
{\it atomic} states defined by the {\it nonlinear} Gross-Pitaevskii equation. 
While the latter are the {\it elementary} excitations corresponding to small 
deviations from a given atomic state and are described by the {\it linear} 
Bogolubov-De Gennes equations [1,2].

Assume that at the initial time the system was condensed to the ground-state 
level with an energy $E_0$. So that the initial condition to Eq. (1) is 
$\varphi({\bf r},0)=\varphi_0({\bf r})$. Suppose we wish to couple the ground 
state $\varphi_0$ with another state $\varphi_j$ having a higher energy 
$E_j$. The best way for doing this is, clearly, by switching on an external 
field, say $\hat V= V({\bf r})\cos\omega t$, oscillating with a frequency
$\omega$ which is close to the transition frequency $\omega_j\equiv
(E_j-E_0)/\hbar$, so that the detuning $\Delta\omega\equiv\omega-\omega_j$ 
be small, $|\Delta\omega/\omega|\ll 1$. Then one can look for a solution to
Eq. (1) in the form
\begin{equation}
\label{2}
\varphi({\bf r},t) = c_0(t)\varphi_0({\bf r})\; e^{-iE_0t/\hbar} +
c_j(t)\varphi_j({\bf r})\; e^{-iE_jt/\hbar} \; .
\end{equation}
The considered situation is very similar to the description of nonlinear 
resonant processes in optics [3,4]. The validity of the quasi-resonant 
two-level approximation (2) for the time-dependent Gross-Pitaevskii equation 
has been confirmed mathematically [5-7] and also proved by direct numerical 
simulations of the Gross-Pitaevskii equation [8,10,12-14], the agreement 
between the two-level picture and the simulations being excellent.

The coefficients $c_i(t)$ define the fractional level populations 
$n_i(t)\equiv |c_i(t)|^2$. The equations for these coefficients can be 
obtained by substituting the presentation (2) in Eq. (1). It is again worth 
noticing the similarity of this presentation with the slowly-varying-amplitude 
approximation commonly used in optics, if one treats the factors $c_i(t)$ as 
slow functions of time, as compared to the exponentials in Eq. (2), which 
implies that $|dc_i/dt|\ll E_i$. This approximation, being complimented by
the averaging technique [20], permits one to slightly simplify the evolution 
equations for $c_i(t)$. In this way [5-7], we come to the equations
\begin{equation}
\label{3}
\frac{dc_0}{dt} =\; - i\alpha\; n_jc_0 -\; \frac{i}{2}\; \beta\; c_j\;
e^{i\Delta\omega t} \; , \qquad
\frac{dc_j}{dt} =\; - i\alpha\; n_0c_j -\; \frac{i}{2}\; \beta^*\; c_0 \;
e^{-i\Delta\omega t}\; ,
\end{equation}
in which the transition amplitudes
$$
\alpha_{ij} \equiv A\;\frac{N}{\hbar}\; \int |\varphi_i({\bf r})|^2
\left ( 2|\varphi_j({\bf r})|^2 -|\varphi_i({\bf r})|^2\right ) \; d{\bf r}\; ,
\qquad
\beta \equiv\; \frac{1}{\hbar}\; \int \varphi_0^*({\bf r}) V({\bf r})
\varphi_j({\bf r})\; d{\bf r} \; ,
$$
caused by the nonlinearity and by the modulating field, respectively, are 
introduced, and the abbreviated notation $\alpha\equiv\alpha_{0j}$, with 
setting $\alpha_{0j}=\alpha_{j0}$, is used. Note that the transition 
amplitude $\beta$ is nonzero only if the potential $V({\bf r})$ depends on 
space variables. The initial conditions to Eqs. (3) are $c_0(0)=1$ and 
$c_j(0)=0$.  

Usually, one solves such evolution equations with fixed parameters $\alpha,\;
\beta$ and $\Delta\omega$, keeping in mind a particular realization. Instead 
of this, we have studied the behavior of solutions to Eqs. (3) in a wide 
range of varying parameters. It turned out that this behaviour is 
surprisingly rich exhibiting new interesting effects.

First of all, it is easy to notice that the number of free parameters in 
Eqs. (3) can be reduced to two by the appropriate scaling. For this purpose, 
we measure time in units of $\alpha^{-1}$ and introduce the dimensionless 
parameters 
$$
b \equiv \frac{|\beta|}{\alpha} \; , \qquad 
\delta \equiv \frac{\Delta \omega}{\alpha} \; .
$$
It is also evident that Eqs. (3) are invariant with respect to the inversion 
$\alpha\rightarrow -\alpha$, $\beta\rightarrow -\beta$, 
$\Delta\omega\rightarrow -\Delta\omega$, and $t\rightarrow -t$. Therefore it 
is possible to fix the sign of one of the parameters, say $\alpha>0$, since 
the opposite case can be obtained by the inversion. For concreteness, we 
shall also keep in mind that $\beta$ is positive. The dimensionless detuning 
is assumed to always be small, $|\delta|\ll 1$. And the dimensionless 
transition amplitude $b$ is varied in the region $0\leq b\leq 1$. We have 
accomplished a careful analysis by numerically solving Eqs. (3). When 
parameter $b$ is small, the fractional populations oscillate reminding 
the Rabi oscillations in optics, where $|\beta|$ would play the role of 
the Rabi frequency. The amplitude of these oscillations increases with 
increasing $b$. It would be more correct to say that in our case there 
exist {\it nonlinear Rabi oscillations}, as far as Eqs. (3) differ from 
the corresponding equations for optical two-level systems by the presence 
of the nonlinearity due to interatomic interactions. This nonlinearity not 
only slightly modifies the Rabi-type oscillations of the fractional 
populations but, for a particular relation between parameters,
can lead to dramatic effects. By accurately analyzing the behaviour of 
solutions to Eqs. (3), with gradually varying parameters, we have found out 
that there exists the {\it critical line} 
$$
b+\delta\simeq 0.5 \; ,
$$ 
at which the system dynamics experiences sharp changes. This is illustrated in 
Figs. 1 to 4, where the parameter $b=0.4999$ is kept fixed and the critical 
line is crossed by varying the detuning $\delta$. In Fig. 1, the detuning is 
zero, and the fractional populations display the Rabi-type oscillations. By 
slightly shifting the detuning to $\delta=0.0001$ drastically changes the 
picture to that in Fig. 2, where the top of $n_j(t)$ and the bottom of 
$n_0(t)$ become flat, and the oscillation period is approximately doubled. 
A tiny further variation of the detuning to $\delta=0.0001001$ yields again 
drastic changes to Fig. 3, where there appear the upward cusps of $n_j(t)$ 
and the downward cusps of $n_0(t)$. The following small increase of the 
detuning to $\delta=0.00011$ squeezes the oscillation period twice, as is 
shown in Fig. 4. After this, making $\delta$ larger does not result in essential
qualitative changes of the population behaviour. All dramatic changes in 
dynamics occur in a tiny vicinity of the critical line. The same phenomena 
happen when crossing the line $b+\delta\simeq 0.5$ at other values of parameters
or if $\delta$ is fixed but $b$ is varied. This is demonstrated in Fig.5 
for varying $\delta$ and another choice of $b=0.3$.

The unusual behaviour of the fractional populations is due to the nonlinearity
of the evolution equations (3). Systems of nonlinear differential equations, 
as is known, can possess qualitatively different solutions for parameters 
differing by infinitesimally small values. The transfer from one type of 
solutions to another type, in the theory of dynamical systems, is, generally, 
termed bifurcation. At a bifurcation line, dynamical system is structurally 
unstable.

The second aim of our paper is to show that the found instability in the
considered dynamical system is analogous to a {\it phase transition} in a
statistical system. To elucidate this analogy for the present case, we have 
to consider the time-averaged features of the dynamical system given by 
Eqs. (3). To this end, we need, first, to define an effective Hamiltonian 
generating the evolution equations. This can be done by transforming 
these equations to the Hamiltonian form 
$$
i\; \frac{dc_0}{dt}=\frac{\partial H_{eff}}{\partial c_0^*}\; , \qquad
i \; \frac{dc_j}{dt} = \frac{\partial H_{eff}}{\partial c_j^*} \; ,
$$ 
with the effective Hamiltonian 
\begin{equation}
\label{4}
H_{eff} = \alpha\; n_0n_j + \frac{1}{2}\left ( \beta\; e^{i\Delta\omega t}
c_0^*c_j + \beta^*\; e^{-i\Delta\omega t} c_j^* c_0\right ) \; .
\end{equation}
An effective energy of the system can be defined as a time average of the 
effective Hamiltonian (4). For this purpose, Eqs. (3) can be treated by 
means of the averaging technique [20], as it is described in detail in 
Ref. [5], which provides the guiding-center solutions. Substituting the 
latter in Eq. (4), together with the time-averaged fractional populations, 
results in the effective energy 
$$
E_{eff} =\frac{\alpha b^2}{2\varepsilon^2} \left ( 
\frac{b^2}{2\varepsilon^2} + \delta \right ) \; , 
$$
where $\varepsilon$ is a dimensionless average frequency defined by the equation 
$$
\varepsilon^4(\varepsilon^2-b^2) =
(\varepsilon^2-b^2-\varepsilon^2\delta)^2 \; .
$$
The effective energy, being the time average of the effective Hamiltonian 
(4), characterizes the average features of the system. As an {\it order 
parameter} for this averaged system, one can take the difference of the 
time-averaged populations, 
$$
\eta \equiv \overline n_0 -\overline n_j=1- \frac{b^2}{\varepsilon^2}\; .
$$
The capacity of the system to store the energy pumped in by the resonant field 
can be described by the {\it pumping capacity} $C_\beta=\partial E_{eff}/
\partial|\beta|$. The influence of the detuning on the order parameter is 
characterized by the {\it detuning susceptibility} $\chi_\delta=
|\partial\eta/\partial\delta|$. Analyzing the behaviour of the introduced 
characteristics as functions of the parameters $b$ and $\delta$, we found 
out that they exhibit critical phenomena at the {\it critical line} $b+\delta=
0.5$, which coincides with the bifurcation line for the dynamical system. 
Expanding these characteristics over the small relative deviation 
$\tau\equiv|b-b_c|/b_c$ from the {\it critical point} $b_c=0.5-\delta$, we 
obtain 
\begin{equation}
\label{5}
\eta-\eta_c\simeq\frac{\sqrt{2}}{2}( 1 - 2\delta)\tau^{1/2}\; , \qquad
C_\beta\simeq\frac{\sqrt{2}}{8}\tau^{-1/2}\; , \qquad
\chi_\delta\simeq\frac{1}{\sqrt{2}}\tau^{-1/2}\; ,
\end{equation} 
where $\eta_c\equiv\eta(b_c)$ and $\tau\rightarrow 0$. As is seen, the 
pumping capacity and detuning susceptibility display divergence at the 
critical point. The related {\it critical indices} for $C_\beta,\; \eta$, 
and $\chi_\delta$ are equal to $1/2$. These indices satisfy the known scaling 
relation: 
$$
ind(C_\beta)+2\; ind(\eta)+ind(\chi_\delta)=2 \; ,
$$
where $ind$ is the evident abbreviation for index.

In order to clarify what is the origin of the found critical effects for 
the studied dynamical system, let us return back to Eqs. (3). Again we pass 
to dimensionless notation measuring time in units of $\alpha^{-1}$. By 
means of the substitution
\begin{equation}
\label{6}
c_0 = \left ( \frac{1-p}{2}\right )^{1/2} \; \exp \left\{ i\left (
q_0 + \frac{\delta}{2}\; t\right ) \right \} \; , \qquad
c_j = \left ( \frac{1+p}{2}\right )^{1/2} \; \exp \left\{ i\left (
q_1 -\; \frac{\delta}{2}\; t\right ) \right \} \; , 
\end{equation}
where $p,\; q_0$, and $q_1$ are real functions of $t$, equations (3) can be 
reduced to the form
\begin{equation}
\label{7}
\frac{dp}{dt} = - b\; \sqrt{1-p^2}\; \sin q \; , \qquad
\frac{dq}{dt} = p + \frac{bp}{\sqrt{1-p^2}}\; \cos q + \delta \; ,
\end{equation}
in which $q\equiv q_1-q_0$. Note that this reduction to an autonomous 
dynamical system is valid for arbitrary detuning $\delta$. Moreover, this 
system possesses the integral of motion 
\begin{equation}
\label{8}
I(p,q) = \frac{1}{2}\; p^2-b \sqrt{1-p^2}\; \cos q + \delta p \;,
\end{equation}
which can be defined by using the initial conditions $p(0)=-1$ and $q(0)=0$ 
corresponding to the conditions $c_0(0)=1$ and $c_j(0)=0$. This gives 
\begin{equation}
\label{9}
I(-1,0)=\frac{1}{2}-\delta \; .
\end{equation}
The existence of the integral of motion means that the dynamical system is 
integrable in quadratures. This fact does not help much for studying the time 
evolution of the system, since the formal solutions $p(t)$ and $q(t)$ are 
expressed through rather complicated integrals, so that the system evolution, 
anyway, is to be analysed numerically. However, the property of integrability 
implies that the appearance of chaos in the system is impossible. Consequently, 
the observed critical effects in no way could be related to chaos. Then what is 
their origin? The answer to this question comes from the analysis of the 
phase portrait for Eqs. (7) in the rectangle defined by the inequalities 
$-1\leq p\leq 1$, $0\leq q\leq  2\pi$. This analysis shows that, if 
$b+\delta<0.5$, then the motion starting at the initial point $p(0)=-1$, 
$q(0)=0$ is oscillatory, with a trajectory lying always in the lower part 
of the phase rectangle, below the separatrix given by the equation 
\begin{equation}
\label{10}
\frac{1}{2} \; p^2-b \sqrt{1-p^2} \; \cos q +\delta p - b =0 \; .
\end{equation}
When $b+\delta=0.5$, the separatrix touches the initial point, so that the 
following motion occurs in the phase region above the separatrix. In this 
way, if we consider, under the given initial conditions, the parametric 
manifold formed by the parameters $b\in[0,1]$ and $\delta\ll 1$, then the 
critical line $b+\delta=0.5$ separates the parametric regions related to 
two different types of solutions to Eqs. (7).

\section{Spatio-Temporal Evolution}

The resonance formation of coherent topological modes can be noticed by 
observing the spatio-temporal behaviour of the atomic density. To 
illustrate this, we consider a cylindrical trap, with the frequency ratio
\begin{equation}
\label{11}
\nu \equiv \frac{\om_z}{\om_r} \qquad (\om_r\equiv\om_x=\om_y) \; .
\end{equation}
Introduce dimensionless cylindrical variables
\begin{equation}
\label{12}
r \equiv \frac{\sqrt{r_x^2 + r_y^2}}{l_r} \; , \qquad z\equiv \frac{r_z}{l_r} 
\qquad \left ( l_r \equiv \sqrt{\frac{\hbar}{m_0\om_r}}\right )
\end{equation}
and the dimensionless wave function
\begin{equation}
\label{13}
\psi_{nmk}(r,\vp,z) \equiv l_r^{3/2}\vp_{nmk}({\bf r}) \; ,
\end{equation}
in which $n=0,1,2,\ldots$ is the radial quantum number; $m=0,\pm 1,\pm 2,\ldots$ 
is the asimuthal quantum number; and $k=0,1,2,\ldots$ is the axial quantum number. 
The effective coupling parameter in dimensionless units is
\begin{equation}
\label{14}
g\equiv \frac{AN}{\hbar\om_r l_r^3}  = 4\pi\;
\frac{a_s}{l_r}\; N \; .
\end{equation}

The density of particles
\begin{equation}
\label{15}
\rho({\bf r},t) \equiv N|\vp({\bf r},t)|^2 \; ,
\end{equation}
according to the form (2), contains both slow functions of time, $c_0(t)$ and
$c_j(t)$, as well as fastly oscillating exponentials. To exclude the fastly
oscillating terms, we introduce the {\it envelope density}
\begin{equation}
\label{16}
\overline\rho({\bf r},t) \equiv \frac{\om}{2\pi}
\int_0^{2\pi/\om} \rho({\bf r},t)\; dt \; ,
\end{equation}
where the integration is over time explicitly entering the exponentials, while
$c_0$ and $c_j$ are kept fixed. Defining the dimensionless envelope density
\begin{equation}
\label{17}
\rho(r,\vp,z,t) \equiv \frac{l_r^3}{N}\; \overline\rho({\bf r},t) \; ,
\end{equation}
we obtain
\begin{equation}
\label{18}
\rho(r,\vp,z,t) = n_0(t)\; |\psi_0(r,\vp,z)|^2 + 
n_j(t)\; |\psi_j(r,\vp,z)|^2 \; , 
\end{equation}
with $j$ denoting the set $\{ n,m,k\}$ of quantum numbers.

Numerical calculations for the envelope density (18) are illustrated in Figs. 
6 to 11, where it is clearly seen how strongly the spatio-temporal behaviour 
of the density (18) depends on the values of the parameters $b$ and $\delta$ 
in the vicinity of the critical line $b+\delta\simeq 0.5$. The behaviour of the 
density for the {\it radial dipole mode} $(n=1,\; m=0,\; k=0)$ is shown in Figs. 
6 and 7. The corresponding transition frequency in units of $\om_r$, for large 
$g\gg 1$, is
$$
\om_{100} \simeq 0.096 (\nu g)^{2/5} \; .
$$
The density for the {\it basic vortex mode} $(n=0,\; m=1,\; k=0)$ is given in
Figs. 8 and 9. The related transition frequency, for $g\gg 1$, is
$$
\om_{010} \simeq \frac{3.424}{(\nu g)^{2/5}} \; .
$$
And the density (18) for the {\it axial dipole mode} $(n=0,\; m=0,\; k=1)$ is
shown in Figs. 10 and 11; the transition frequency of the latter mode being
$$
\om_{001} \simeq 0.060 (\nu g)^{2/5} \; .
$$
In all Figs. 6 to 11, we set the coupling parameter (14) g=100 and take the
frequency ratio (11) $\nu=10$, which corresponds to a disk-shape trap with the
aspect ratio
\begin{equation}
\label{19}
R_r \equiv \left ( \frac{<r^2>_{nmk}}{<z^2>_{nmk}}\right )^{1/2} \simeq
\sqrt{2}\; \nu \; ,
\end{equation}
as $g\gg 1$. For the chosen values of $g$ and $\nu$ the transition frequencies 
are
$$
\om_{010}\simeq 0.216 \qquad (basic \; vortex\; mode) \; ,
$$
$$
\om_{001}\simeq 0.951 \qquad (axial \; dipole\; mode) \; ,
$$
$$
\om_{100}\simeq 1.521 \qquad (radial \; dipole\; mode) \; .
$$
This demonstrates that the transition frequencies are arranged in the order
\begin{equation}
\label{20}
\om_{010} < \om_{001} < \om_{100} \; ,
\end{equation}
the lowest energy level corresponding to the basic vortex mode.

Let us note that the basic vortex mode, with the winding number $|m|=1$, sets 
off from other vortices with higher winding numbers $|m|\geq 2$. For a given
angular momentum, the creation of several basic vortices is more energetically
profitable than the formation of one vortex with a higher winding number 
[2,16,18].

\section{Density of Current}

When a coherent topological mode is excited in a trap, the density of current, 
as a function of space and time, also displays a peculiar behaviour. The density
of current
\begin{equation}
\label{21}
{\bf j}({\bf r},t) \equiv -\; \frac{i\hbar}{2m_0} \left [
\vp^*({\bf r},t){\bf\nabla}\vp ({\bf r},t) - \vp({\bf r},t)
{\bf\nabla}\vp^*({\bf r},t) \right ]
\end{equation}
can be rewritten as
\begin{equation}
\label{22}
{\bf j}({\bf r},t) =\frac{\hbar}{m_0}\; {\rm Im}\; \vp^*({\bf r},t)
{\bf\nabla} \vp({\bf r},t) \; .
\end{equation}
The wave function (2) is the sum
\begin{equation}
\label{23}
\vp({\bf r},t) =\vp_0({\bf r},t) + \vp_j({\bf r},t)
\end{equation}
of the terms
\begin{equation}
\label{24}
\vp_i({\bf r},t) \equiv c_i(t) \vp_i({\bf r}) e^{-iE_i\;t/\hbar} \; ,
\end{equation}
where the index $i=0,\; j$.

With the wave function (23), the density of current (21) becomes
\begin{equation}
\label{25}
{\bf j}({\bf r},t) = {\bf j}_0({\bf r},t)  + {\bf j}_{top}({\bf r},t)
+ {\bf j}_{int}({\bf r},t) \; ,
\end{equation}
where the first term is the {\it ground-state current density}
\begin{equation}
\label{26}
{\bf j}_0({\bf r},t) \equiv \frac{\hbar}{m_0} \; {\rm Im}\;
\vp_0^*({\bf r},t) {\bf\nabla}\vp_0({\bf r},t) \; ,
\end{equation}
the second term is the {\it topological current density}
\begin{equation}
\label{27}
{\bf j}_{top}({\bf r},t) \equiv \frac{\hbar}{m_0} \; {\rm Im}\;
\vp_j^*({\bf r},t) {\bf\nabla}\vp_j({\bf r},t) \; ,
\end{equation}
due to the excited topological mode, and the third term is the {\it
interference current density}
\begin{equation}
\label{28}
{\bf j}_{int}({\bf r},t) \equiv \frac{\hbar}{m_0} {\rm Im}
\left [ \vp_0^*({\bf r},t){\bf\nabla}\vp_j({\bf r},t) +
\vp_j^*({\bf r},t){\bf\nabla}\vp_0({\bf r},t) \right ] \; ,
\end{equation}
caused by the interference between the ground-state mode and the excited 
topological mode.

For the functions (24), one has
$$
\vp_i^*({\bf r},t){\bf\nabla}\vp_i({\bf r},t)  =
|c_i(t)|^2 \vp_i^*({\bf r}){\bf\nabla}\vp_i({\bf r}) \; .
$$
Using the substitution (6), one gets
\begin{equation}
\label{29}
n_0(t) \equiv |c_0(t)|^2 = \frac{1-p(t)}{2} \; , \qquad
n_j(t) \equiv |c_j(t)|^2 = \frac{1+p(t)}{2} \; .
\end{equation}
Since the ground-state wave function $\vp_0({\bf r})$ is real,
\begin{equation}
\label{30}
{\bf j}_0({\bf r},t) = 0 \; .
\end{equation}
And the topological current density (27) acquires the form
\begin{equation}
\label{31}
{\bf j}_{top}({\bf r},t) = \frac{\hbar}{m_0}\; n_j(t)
{\rm Im} \vp_j^*({\bf r}){\bf\nabla} \vp_j({\bf r}) \; .
\end{equation}
For example, in the case of a vortex mode, when $\vp_j({\bf r})\sim e^{im\vp}$, 
with $m=0,\pm 1,\pm 2,\ldots$, one has
$$
{\rm Im} \vp_j^*({\bf r}) {\bf\nabla}\vp_j({\bf r}) = 
\frac{m}{r}\; |\vp_j({\bf r})|^2 {\bf e}_\vp \; .
$$
Then the topological current density (31) is
\begin{equation}
\label{32}
{\bf j}_{top}({\bf r},t) = \frac{m\hbar}{m_0 r}\; n_j(t) |\vp_j({\bf r})|^2\;
{\bf e}_\vp \; .
\end{equation}
The total topological current is, of course, zero,
\begin{equation}
\label{33}
\int {\bf j}_{top}({\bf r},t)\; d{\bf r} = 0 \; ,
\end{equation}
provided there is no flux through the boundary.

The interference current density (28) is a kind of the Josephson current
oscillating with the transition frequency $\om_j\approx\om$. If one averages 
over these fast oscillations, treating $c_i(t)$ as slow functions of time, one 
gets
\begin{equation}
\label{34}
\frac{\om}{2\pi} \int_0^{2\pi/\om} {\bf j}_{int}({\bf r},t) \; dt = 0 \; .
\end{equation}
Thus, the slow time variation of the summary density of current (25) is
\begin{equation}
\label{35}
\frac{\om}{2\pi} \int_0^{2\pi/\om} {\bf j}({\bf r},t) \; dt = 
{\bf j}_{top}({\bf r},t) \; ,
\end{equation}
being due to the topological current density (31).

\section{Discussion}

Here we have studied the resonant excitation of coherent topological modes 
in a trapped Bose-Einstein condensate. It is worth mentioning that there 
could be another way of creating those stationary modes that do not have 
zeros, except, may be, at infinity. Suppose, we wish to create a mode 
$f({\bf r})$ having no zeros. Then it is possible to perturb the system 
with the potential
$$
V_f({\bf r}) = E_f -\; \frac{\hat H(f)f({\bf r})}{f({\bf r})} \; ,
$$
such that the given function $f({\bf r})$ be the ground-state wave function
for the eigenproblem
$$
\left [ \hat H(f) + V_f({\bf r})\right ] \; f({\bf r}) = E_f f({\bf r}) \; .
$$
However, this way does not allow us to form topologically different modes
having different number of zeros, which is admissible by means of the resonant
excitation.

In conclusion, we have considered the population dynamics of a 
trapped Bose-Einstein condensate, subject to the action of a resonant 
spatio-temporal modulation of the trapping potential. A careful 
analysis of evolution equations has been made for the wide range 
of varying parameters. Such a variation of parameters can be easily 
realized by changing trap characteristics, varying the number and 
kind of condensed atoms, and by changing the scattering length
using Feshbach resonances [1,2]. It turned out that on the manifold 
of possible parameters there exists a critical line where the evolution 
equations display structural instability. We have demonstrated that this
instability for a dynamical system is analogous to a phase transition 
for a stationary averaged system. For the latter, one can define an order 
parameter, pumping capacity, and detuning susceptibility which exhibit 
critical phenomena at the critical line. The origin of this critical line 
is elucidated by showing that it divides the parametric manifold onto two 
regions corresponding to different solutions of the evolution equations.
The spatio-temporal behaviour of the density of trapped atoms is studied 
for several first topological modes.

\vskip 5mm

{\bf Acknowledgement}

\vskip 3mm

One of us (V.I.Y.) is very grateful to V.K. Melnikov  for many useful 
discussions.

\newpage

\begin{center}
{\bf Figure captions}
\end{center}

{\bf Fig. 1.} The time dependence of the fractional populations $n_0(t)$ and
$n_j(t)$ for $b=0.4999$ and $\delta=0$. Here and in the Figures 1 to 5
the dashed line corresponds to the ground-state population $n_0(t)$ and the
solid line, to the excited-level population $n_j(t)$.

\vskip 5mm

{\bf Fig. 2.} Flattening of the fractional populations, with their
oscillation period being doubled, at $b=0.4999$ and $\delta=0.0001$.

\vskip 5mm

{\bf Fig. 3.} The appearance of the upward cusps of $n_j(t)$ and of the
downward cusps of $n_0(t)$ for $b=0.4999$ and $\delta=0.0001001$.

\vskip 5mm

{\bf Fig. 4.} Fractional populations versus time for $b=0.4999$ and
$\delta=0.00011$.

\vskip 5mm

{\bf Fig. 5.} The dynamics of fractional populations, for varying $\delta$
and fixed $b=0.3$, demonstrating the qualitative change of behaviour when
crossing the critical line: (a) $\delta=0.242681$; (b) $\delta=0.242682$;
(c) $\delta=0.243000$; (d) $\delta=0.3$.

\vskip 5mm

{\bf Fig. 6.} The spatio-temporal behaviour of the envelope density (18) 
as a function of the radial variable $r$, at fixed $z=0$ for the radial 
dipole mode $(n=1,\; m=0,\; k=0)$. Here and in all following figures the 
characteristic parameters $g=100,\; \nu=10$, and $b=0.4999$ are fixed,
while the detuning $\delta$ is varied in the vicinity of the critical line
$b+\delta\simeq 0.5$. In this Figure, we set $\delta=0.00010$. Time is
measured in dimensional units explained in the text: $t=0$ (solid line);
$t=3$ (long-dashed line); $t=15$ (short-dashed line).

\vskip 5mm

{\bf Fig. 7.} The same as in Fig. 6, but for the detuning $\delta=0.00011$
and for the moments of time: $t=0$ (solid line); $t=10$ (long-dashed line);
$t=26$ (short-dashed line). At the time $t=26$ the system is in the pure
radial dipole state.

\vskip 5mm

{\bf Fig. 8.} Excitation of the basic vortex mode $(n=0,\; m=1,\; k=0)$. 
The envelope density is shown as a function of the radial variable $r$
for $z=0$. The detuning is $\delta=0.00010$. The moments of time are:
$t=0$ (solid line); $t=3$ (long-dashed line); $t=15$ (short-dashed line).

\vskip 5mm

{\bf Fig. 9.} The same as in Fig. 8, but for the detuning $\delta=0.00011$.
The moments of time are: $t=0$ (solid line); $t=10$ (long-dashed line); 
$t=26$ (short-dashed line). At $t=26$ the system is in the pure
vortex state.

\vskip 5mm

{\bf Fig. 10.} Excitation of the axial dipole mode $(n=0,\; m=0,\; k=1)$. 
The envelope density as a function of the axial variable $z$ at the point 
$r=0$. Here $\delta=0.00010$. The time moments are: $t=0$ (solid line); 
$t=3$ (long-dashed line); $t=15$ (short-dashed line).

\vskip 5mm

{\bf Fig. 11.} The same as in Fig. 10, but for the detuning $\delta=0.00011$
and for the times: $t=0$ (solid line); $t=20$ (long-dashed line); $t=26$ 
(short-dashed line). At the moment $t=26$ the system is in the pure axial 
dipole state.

\end{document}